\documentstyle[twocolumn]{jpsj}
\title{Resonant X-Ray Scattering from the Quadrupolar Ordering Phase of CeB$_6$}

\author{Tatsuya {\sc Nagao}\footnote{E-mail: tnagao@phys.sci.gunma-u.ac.jp}
    and Jun-ichi {\sc Igarashi}}

\inst{Faculty of Engineering, Gunma University, Kiryu, Gunma 376-8515}

\recdate{ \hspace*{3.0cm} }

\kword{resonant x-ray scattering, CeB$_{6}$, 
       orbital ordering, Ce $L_{{\rm III}}$ absorption edge}

\abst{
We theoretically investigate the origin of the resonant x-ray  scattering 
(RXS) signal near the Ce $L_{\rm III}$ absorption edge in the quadrupolar 
ordering phase of CeB$_6$, considering the intersite interaction between 
the $\Gamma_8$ states in the initial state.
The anisotropic charge distribution of the $4f$ states modulates
the $5d$ states through the intra-atomic Coulomb interaction
and thereby generates a large RXS superlattice intensity.
The temperature and magnetic field dependence indicates that
the induced dipolar and octupolar orders have
little influence on the RXS spectra,
in good agreement with the recent experiment. 
}

\begin{document}
\sloppy
\maketitle
\input epsf.sty


Resonant x-ray scattering (RXS) has attracted much interest
as a useful tool to investigate the orbital order,
since the resonant enhancement on the superlattice Bragg spots 
for the orbital order was observed near the Mn $K$ edge 
in LaMnO$_3$.~\cite{Murakami}
Although similar experiments have already been carried out on several
materials, the origin of the RXS signal is still controversial.
The RXS is described by a second order process 
that photons are virtually absorbed by exciting a core electron 
to unoccupied states and then emitted by recombining the excited electron 
with the core hole. Therefore, the $4p$ states are involved near the $K$ edge 
for LaMnO$_3$ in the dipolar process.
In the early stage, the Coulomb interaction between the $3d$ and $4p$ states
was considered as the origin,~\cite{Ishihara1,Ishihara2}
but subsequent studies based on the band structure calculations
\cite{Elfimov,Benfatto,Takahashi1,Takahashi2} 
have revealed that the Coulomb effect is two order of magnitude smaller
than the effect of the Jahn-Teller distortion,
indicating that the $4p$ states are sensitive to electronic structure 
of neighboring sites. 

Recently the RXS experiment was carried out near the Ce $L_{\rm III}$ 
absorption edge in the quadrupolar ordering phase of CeB$_6$.~\cite{Nakao}
The lattice distortion is not observed because of much weaker coupling
between lattice and well-localized $4f$ states.
Also, the $5d$ states are considered to be more localized than 
the $4p$ states in LaMnO$_3$.
Therefore, the mechanism of the RXS in CeB$_6$ is expected to be different
from that in LaMnO$_3$.
The purpose of this paper is to make clear the origin and to discuss
the implication of the RXS spectra in comparison with the experiment,
by considering the intersite interaction between the $\Gamma_8$ states.

In CeB$_6$, each Ce atom is considered to be in the $f^1$-configuration.
Within the $J=5/2$ subspace, the $\Gamma_8$ quartet states have 
lower energy than the $\Gamma_7$ doublet under the cubic crystal field.
The $\Gamma_8$-$\Gamma_7$ excitation energy is $\sim 540$ K.
With decreasing temperatures the antiferro quadrupole (AFQ) order appears
at $T_Q= 3.3$ K with an ordering 
wave vector ${\rm Q}=(1/2,1/2,1/2)$. 
Antiferromagnetic phases exist at even lower temperatures,
but will not be considered in this paper.
These phase transitions originate from the intersite interaction,
which lifts the degeneracy of the $\Gamma_8$ states.

The intersite interaction was derived on the basis of a RKKY interaction 
by Ohkawa.~\cite{Ohkawa1,Ohkawa2}
Recently, Shiina {\it et} {\it al}.~\cite{Shiina,Sakai,Shiba} extended his model 
by fully taking account of the symmetry of the interaction 
as well as the order parameters. Thereby, they solved a longstanding 
controversy between the neutron diffraction\cite{Effantin} and 
NMR\cite{Takigawa} in the context of the induced order parameters
under the external magnetic field. 
Following their study, we use in the initial state the Hamiltonian 
$\tilde H$ of the so called {\em extended Ohkawa model}:~\cite{Shiba}
\begin{eqnarray}
& &\tilde H = D \sum_{\langle i,j \rangle}
\left[ (1+\delta)
  \mbox{\boldmath$\mu$}_{i} \cdot \mbox{\boldmath$\mu$}_{j}
  + \tau_{i}^{y} \tau_{j}^{y}
  + \epsilon \mbox{\boldmath$\sigma$}_{i} \cdot \mbox{\boldmath$\sigma$}_{j}
\right.  \nonumber \\
& & \hspace*{0.80cm} \left.
  + \frac{1+\epsilon}{2} 
   (  \mbox{\boldmath$\tau$}_{i} ' \cdot \mbox{\boldmath$\tau$}_{j} '
    + \mbox{\boldmath$\eta$}_{i} \cdot \mbox{\boldmath$\eta$}_{j}
    + \mbox{\boldmath$\zeta$}_{i} \cdot \mbox{\boldmath$\zeta$}_{j}
    ) \right] \nonumber \\
    & & \hspace*{1.0cm} - 2 \mu_{\rm B} 
   \sum_{i} \left( \mbox{\boldmath$\sigma$}_{i} 
                 + \frac{4}{7} \mbox{\boldmath$\eta$}_{i} \right)
        \cdot {\bf H},
\label{eq.ohkawa}
\end{eqnarray}
where 
$\mbox{\boldmath$\mu$}$, $\mbox{\boldmath$\tau$} '$,
$\mbox{\boldmath$\eta$}$, $\mbox{\boldmath$\zeta$}$ 
acting on the $\Gamma_8$ states
are explicitly defined in ref. 13.
Since the RXS spectra are not sensitive to the parameter values of $\delta$ 
and $\epsilon$, we simply assume $\delta=0.2$ and $\epsilon=1$.
This choice is known to give a reasonable 
description for the AFQ phase.~\cite{Shiina,Sakai,Shiba}
Assuming $D=1.83$ K, we get the transition temperature $T_Q=3.3$ K 
for the AFQ phase with no external magnetic field in the mean field 
approximation. 

Near the Ce $L_{\rm III}$ absorption edge, the $2p$ core electron is virtually
excited to the $5d$ states and is recombined with the core hole
in the dipolar process.
Since the $2p$ states are well localized around Ce sites,
the scattering tensor can be approximated by a sum of the contributions 
from each site of the created core hole.
The corresponding cross section is given by
\begin{equation}
 \left. \frac{d\sigma}{d\Omega}\right|_{\mu\to\mu'} \propto
 \left| \sum_{\alpha\alpha'}P'^{\mu'}_{\alpha}
    M_{\alpha\alpha'}({\bf G},\omega)P^{\mu}_{\alpha'}
  \right|^2 ,
\end{equation}
with 
\begin{eqnarray}
& & M_{\alpha\alpha'}({\bf G},\omega) = \frac{1}{\sqrt{N}}
  \sum_j\sum_{n,\Lambda}p_n(j) \nonumber \\
& \times &  \frac{\langle\psi_n(j)|x_\alpha(j)|\Lambda\rangle
  \langle \Lambda|x_{\alpha'}(j)|\psi_n(j)\rangle}
       {\hbar\omega-(E_{\Lambda}-E_n(j))+i\Gamma} \exp(i{\bf G}\cdot{\bf r}_j).
\label{eq.dipole}
\end{eqnarray}
Here ${\bf G}$ is the scattering vector,
and $\omega$ is the frequency of photon. 
The $j$ runs over Ce sites.
The cross section becomes order $N$, the number of Ce sites.
The dipole operators $x_\alpha(j)$'s are defined as
$x_1(j)=x$, $x_2(j)=y$, and $x_3(j)=z$ in the coordinate frame fixed 
to the crystal axes with the origin located at the center of site $j$.
The energy eigenvalues are defined as $E_n(j)$ ($n=1\sim 4$)
and the eigenfunctions as $|\psi_n(j)\rangle$ at site $j$ in the mean field
approximation.
We take thermal average on these states with probability $p_n(j)$.
State $|\Lambda\rangle$ represents the intermediate state
with energy $E_{\Lambda}$.
The life-time broadening width $\Gamma$ of the core hole is assumed to be 2 eV.
We study only the situation of ${\bf G}=(1/2,1/2,1/2)$ in the following.
For this ${\bf G}$ value in the conventional scattering geometry,~\cite{Murakami}
the geometrical factors $P^\mu$ for incident 
photons and $P'^{\mu '}$ for scattered photons are given by\cite{Igarashi}
$(P^\sigma)_1=(P'^{\sigma '})_1=(\cos\beta\cos\psi+\sin\psi)/\sqrt{2}$,
$(P^\sigma)_2=(P'^{\sigma '})_2=(\cos\beta\cos\psi-\sin\psi)/\sqrt{2}$,
$(P^\sigma)_3=(P'^{\sigma '})_3=-\sin\beta\cos\psi$,
$(P'^{\pi '})_1=[-\sin\theta(\cos\beta\sin\psi-\cos\psi)+\cos\theta\sin\beta]
/\sqrt{2}$,
$(P'^{\pi '})_2=[-\sin\theta(\cos\beta\sin\psi+\cos\psi)+\cos\theta\sin\beta]
/\sqrt{2}$,
$(P'^{\pi '})_3=[\sin\theta\sin\beta\sin\psi+\cos\theta\cos\beta]$,
with $\theta$ being the Bragg angle and $\beta=\arccos{1/\sqrt{3}}$.
The azimuthal angle $\psi$ is defined such that the crystal axis 
$(1,-1,0)$ lies on the scattering plane at $\psi=0$.

In the intermediate state, the $5d$ states of Ce are involved.
From the band calculation for LaB$_6$,~\cite{Harima}
it is known that the $5d$ states of La hybridize with B
$2p$ states, forming energy bands with their width $\sim 5$ eV. 
The $5d$ states in CeB$_6$ are considered to be similar to those in LaB$_6$, 
except for the relative position to the $4f$ levels.
Note that the $5d$ states are more localized than the $4p$ states
in LaMnO$_3$, since the latter band width is as large as $\sim 20$ eV.
Therefore, the dominant contribution to the RXS spectra
comes from the local process which can be expressed
in terms of the {\em local} Green's function constructed by
the $5d$ density of states (DOS).
We roughly simulate it by a simple rectangular shape 
with the width 5 (2) eV and the edge located 2.5 (5.5) eV above the $4f$ 
levels, for the $t_{2g}$ ($e_g$) symmetry.
The $5d$ band is slightly filled by electrons, which mediate the intersite
interaction mentioned above.~\cite{Ohkawa1,Ohkawa2,Shiba}
In the present calculation, however, we simply assume the vacant $5d$ band 
in the initial state, since the actual filling modifies little
the RXS spectra.
Thus, the local Green's function for the $5d$ states is defined by
\begin{equation}
 G^{5d}_{m^d,m'^d}(\omega)
   = \delta_{m^d m'^d}
     \int \frac{\rho^{5d}_{m^d}(\epsilon)}{\omega-\epsilon+i\delta}
     {\rm d}{\epsilon},
\end{equation}
where $\rho^{5d}_{m^d}(\epsilon)$ is the DOS mentioned above
with $m^d$ indicating either $t_{2g}$ or $e_g$ states.

The intermediate state at {\em the site of core hole} are specified 
by $m^ds^d$ ($s^d$: spin) of the $5d$ electron,
$m^fs^f$ ($m^f=-3,\dots,3$) of the $4f$ electron, and
$m^ps^p$ ($m^p=-1,0,1$) of the $2p$ hole.
Then the resolvent in the space of these states is given by
\begin{eqnarray}
 & & \left(\frac{1}{\omega - H_{\rm int} + i\delta}\right)_
{m^ds^dm^fs^fm^ps^p;m'^ds'^dm'^fs'^fm'^ps'^p} \nonumber \\
 &  = &[G^{5d}_{m^d,m'^d}(\omega+i\Gamma-\epsilon_f-\epsilon_p)^{-1} 
  \nonumber\\ 
  &  &- V_{m^ds^dm^fs^fm^ps^p;m'^ds'^dm'^fs'^fm'^ps'^p}]^{-1},
\label{eq.matrix}
\end{eqnarray}
where $H_{\rm int}$ represents the Hamiltonian in the intermediate state.
The $\epsilon_f$ and $\epsilon_p$ represent the average energy of 
the $4f$ electron and $2p$ hole.
The interaction $V$ contains the spin-orbit interaction for each particles,
and the intra-atomic Coulomb interaction between them.
The parameter values for them are evaluated in the $2p^54f^15d^1$ configuration 
of the Ce$^{3+}$ atom within the Hartree-Fock approximation.~\cite{Cowan}
It is known that the Slater integrals $F^0(5d,4f)$, $F^0(2p,4f)$ 
and $F^0(2p,5d)$ are well screened while the other integrals are not
much even in solids.
We assume relatively small values, 4 eV, 12 eV, and 7 eV
for $F^0(5d,4f)$, $F^0(2p,4f)$ and $F^0(2p,5d)$, respectively,
while we use the atomic values multiplied by a factor 0.8 
for other quantities.  Note that the anisotropic parts
$F^{2}(5d,4f), F^{4}(5d,4f), G^{1}(5d,4f), G^{3}(5d,4f), G^{5}(5d,4f)$
in the Coulomb interaction between the 5d and 4f electrons, not the
isotropic part $F^{0}(5d,4f)$, are relevant to giving rise to the
RXS intensity.
The crystal field energy, which is important for selecting the $\Gamma_8$ 
states as the initial state, is negligibly small in the calculation of 
the intermediate state.
The right hand side of eq.~(\ref{eq.matrix}) is a matrix 
with dimensions $840\times 840$, which we numerically invert.
This resolvent is the same at all sites of core hole,
so that the scattering amplitudes become different at different
sublattices of the AFQ order only after multiplying the matrix elements 
of the dipole operators between the initial and the intermediate state
(see eq.~(\ref{eq.dipole})).

\begin{figure}
\centerline{\epsfxsize=8.0cm\epsfbox{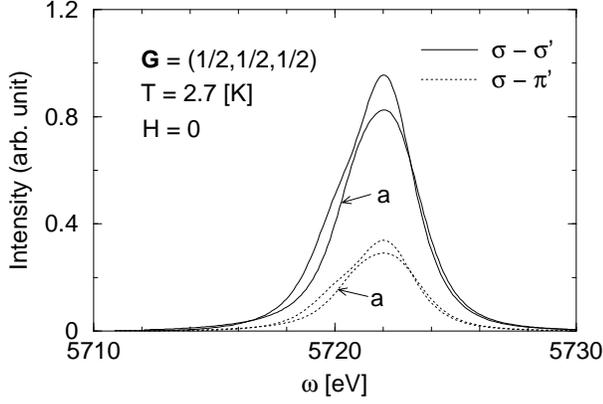}}
\caption{
RXS spectra as a function of photon energy for ${\bf G}=(1/2,1/2,1/2)$.
$H=0$, $T=2.7$ K, and $\psi=0$.
The contributions from three domains are averaged.
The solid and dotted lines represent the intensities 
for the $\sigma\to\sigma'$ channel and 
the $\sigma\to\pi'$ channel, respectively.
Curves with letter ``a" represent the spectra calculated by turning off
the core hole potential.
\label{fig.spec}
}
\end{figure}

Now we discuss the calculated results.
We have three types of possible ordered phase for $H=0$, 
in which one of the staggered quadrupole moments, $\langle O_{xy}\rangle$
($\equiv 4\langle\tau_y\sigma_z\rangle$), 
$\langle O_{yz}\rangle$
($\equiv 4\langle\tau_y\sigma_x\rangle$),
and $\langle O_{zx}\rangle$
($\equiv 4\langle\tau_y\sigma_y\rangle$), 
is finite.
Here $\langle X\rangle$ indicates the thermal average of
operator $X$.
We simply call them as the $O_{xy}$, $O_{yz}$, and $O_{zx}$ phases.
Assuming three domains with equal relative volumes,
we average the RXS intensities over the domains.
Figure \ref{fig.spec} shows the RXS spectrum as a function of photon energy,
at $T=2.7$ K and $\psi=0$.  We obtain a single peak structure in agreement 
with the experiment. ~\cite{Nakao}
 We adjusted the core hole energy $\epsilon_p$ such that 
the peak is located at the experimental position $\hbar\omega=5722$ eV.
The $\omega$-dependence is the same in the three domains.
We roughly estimate that the peak intensity is less than 10$^{-2}$ or 10$^{-3}$ of
that for LaMnO$_3$.~\cite{Takahashi1,Takahashi2}
The contribution of the quadrupolar process is included 
in the curve.~\cite{Com1}
Although it is expected to give a peak in the pre-edge region, 
it is too small to be seen.
This contrasts with the RXS spectra from the AFQ phase
of DyB$_2$C$_2$, where a pre-edge peak is conspicuous.~\cite{Hirota,Tanaka}
The core hole potential cannot become a main origin for the RXS spectra,
since it works the same way at all sites.
This point is clarified by comparing the spectra with those
calculated by turning off the potential.
The latter also consists of a single peak located at the position
about 19 eV higher than the former.
In the figure, the spectrum is shifted such that the peak is 
located at the same position; the spectral shape and intensity are similar 
except for the energy shift. 
This behavior is quite different from the absorption spectra, 
which are known to be considerably enhanced by the core hole potential.

We obtain the scattering amplitudes 
$M({\bf G},\omega)$ in simple forms,
\begin{equation}
    \left( \begin{array}{ccc}
          0 & a & 0 \\
          a & 0 & 0 \\
          0 & 0 & 0 
       \end{array}
\right), 
\left( \begin{array}{ccc}
          0 & 0 & 0 \\
          0 & 0 & a \\
          0 & a & 0 
       \end{array}
\right), 
\left( \begin{array}{ccc}
          0 & 0 & a \\
          0 & 0 & 0 \\
          a & 0 & 0 
       \end{array}
\right) 
\end{equation}
in the $O_{xy}$, $O_{yz}$, and $O_{zx}$ phases, respectively.
Therefore, the principal axes, for example in the $O_{xy}$ phase,
are given by rotating the $x$, $y$ axes with $\pi/4$ around the $z$ axis.
Note that the eigenstates with the eigenvalue 1 (-1) for the operator
$O_{xy}$ has a larger charge distribution along the $(1,-1,0)$ ($(1,1,0)$)
direction.  The principal axes are consistent with the anisotropic
charge density of the $4f$ states in the AFQ phase.
This contrasts with the situation of LaMnO$_3$, where the neighboring
oxygen potentials are more important through
the Jahn-Teller distortion.~\cite{Elfimov,Benfatto,Takahashi1,Takahashi2}

\begin{figure}
\centerline{\epsfxsize=8.0cm\epsfbox{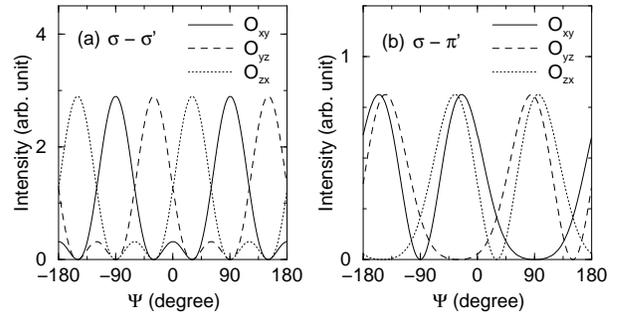}}
\caption{
\noindent
Azimuthal angle dependence of the RXS intensity of the peak 
at $\hbar\omega=5722$ eV, for (a) the $\sigma\to\sigma'$ channel 
and (b) the $\sigma\to\pi'$ channel.
$T=2.7$ K and $H=0$.
The solid, broken, and dotted lines represent the contributions
from the $O_{xy}$, $O_{yz}$, and $O_{zx}$ phases, respectively.
\label{fig.azim}
}
\end{figure}

Figure \ref{fig.azim} shows the azimuthal angle dependence of
the intensity of the peak at $\hbar\omega=5722$ eV.
The contributions from the three domains are separately shown.
The dependence for the $\sigma\to\sigma'$ channel is quite different 
from that for the $\sigma\to\pi'$ channel.
The curves for the $O_{xy}$, $O_{yz}$ and $O_{zx}$ phases can be 
transformed into those for the $O_{yz}$, $O_{zx}$, and $O_{xy}$ phases,
respectively, by shifting $\psi$ with $2\pi/3$.
This threefold symmetry around the $(1,1,1)$ direction perfectly matches
the relation between the order parameters of the three domains.

\begin{figure}
\centerline{\epsfxsize=8.0cm\epsfbox{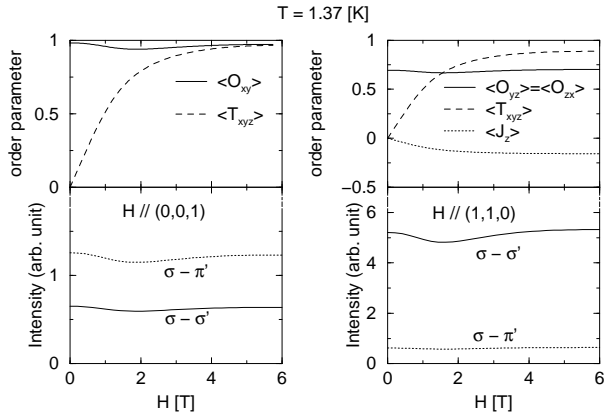}}
\caption{
RXS intensity of the peak at $\hbar\omega=5722$ eV
as a function of $H$.
Left panel: $H\parallel (001)$; Right panel: $H\parallel (110)$.
$T=1.37$ K and $\psi=0$.
The solid and dotted lines represent the intensities for
the $\sigma\to\sigma'$ channel and the $\sigma\to\pi'$ channel,
respectively.
Upper panels show the staggered quadrupole moments,
$\langle O_{xy}\rangle$, $\langle O_{yz}\rangle$, $\langle O_{zx}\rangle$,
the staggered dipole moments $\langle J_z\rangle$, 
and the staggered octupole moment $\langle T_{xyz}\rangle$.
\label{fig.external}
}
\end{figure}

Under the external magnetic field, the staggered octupole moment
and the staggered dipole moment are induced in addition to the
staggered quadrupole moment. To make clear the effects of such induced
moments on the RXS spectra, we calculate the RXS intensity of the peak 
at $\hbar\omega=5722$ eV as a function of $H$.
The temperature is set rather low, $T=1.37$ K, so that the quadrupole order
parameter is sufficiently developed for all values of $H$ 
(the possible antiferromagnetic phase is disregarded).
Figure \ref{fig.external} shows the calculated results in comparison with
the order parameters.
For $H\parallel (001)$, the primary order parameter is the staggered
quadrupole moment of $\langle O_{xy}\rangle$. 
The staggered octupole moment $\langle T_{xyz}\rangle$ 
($\equiv 2\langle\tau^y\rangle$) is induced with increasing $H$,
but the staggered dipole moment is not.
As shown in the left panel, the RXS intensity changes little with changing $H$, 
indicating that the induced octupole moment is irrelevant to the RXS spectra.
For $H\parallel (110)$, the primary order parameter is the staggered
quadrupole moment $\langle O_{yz}+O_{zx}\rangle/\sqrt{2}$,
and the staggered dipole moment $\langle J_z\rangle$ 
($\equiv\langle \frac{14}{11}\sigma^z+\frac{16}{11}\tau^z\sigma^z\rangle$)
is induced in addition to the induced octupole moment with increasing $H$. 
As shown in the right panel, the RXS intensity changes little with
increasing H, indicating that the induced dipole moment is also
irrelevant to the RXS spectra.
This marks the difference between the RXS and the neutron scattering 
experiment, the latter of which detects 
the induced dipole moment.~\cite{Effantin}

\begin{figure}
\centerline{\epsfxsize=8.0cm\epsfbox{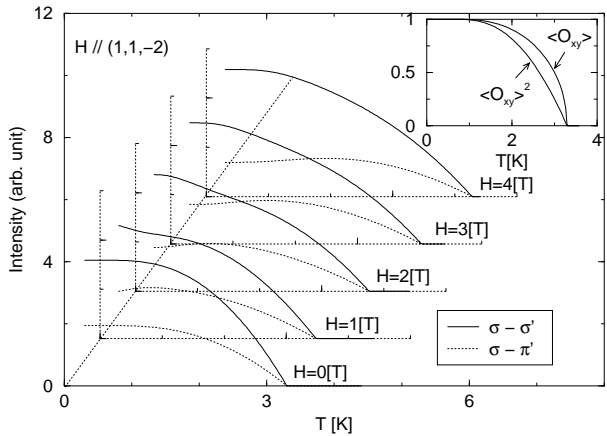}}
\caption{
Temperature dependence of the RXS intensity of the peak 
at $\hbar\omega=5722$ eV, for $H=0$ and for finite $H\parallel (1,1,-2)$. 
The curve for $H=0$ is not the limit of $H\to 0$ but the domain average.
Inset: the staggered quadrupole moment and its square for $H=0$.
\label{fig.phase}
}
\end{figure}

Figure \ref{fig.phase} shows the temperature dependence of the RXS intensity 
of the peak at $\hbar\omega=5722$ eV, for $H=0$ and for several values 
of $H\parallel(1,1,-2)$. 
The inset shows the staggered order parameter ($\langle O_{xy}\rangle$
in the $O_{xy}$ phase) for $H=0$ and its square, calculated within 
the mean field approximation.
The temperature dependence of the spectra for $H=0$ looks
similar to that of the square of the order parameter.
The dependence changes little with changing values of $H$, 
in agreement with the recent experiment.~\cite{Nakao}

In summary, we have calculated the RXS spectra near the Ce $L_{\rm III}$
absorption edge, considering the intersite interaction between the
$\Gamma_8$ states in the initial state.
We have demonstrated that the $4f$ charge distribution modulates
the $5d$ states through the intra-atomic Coulomb interaction 
and thereby generates a large RXS superlattice intensity.
This mechanism is different from that in LaMnO$_3$, where the oxygen
potentials at neighboring sites are much important.
We have evaluated the temperature and magnetic field
dependence of the RXS intensity,
in good agreement with the recent experiment.
The magnetic field dependence indicates 
that the induced dipolar and octupolar orders have
little influence on the RXS intensity.
Thus, it can be concluded that the RXS spectrum in CeB$_6$ is a direct 
reflection of the AFQ order, in contrast with the neutron diffraction
and NMR experiments.
We have used a simplified model for the 5d bands.
A change in the 5d DOS shape and the filling actually modifies
the spectral intensity, but does not alter the qualitative features 
discussed in this paper, such as a single-peak
shape, the temperature and the magnetic field dependences.
A quantitative study using a more realistic model is left for future.

We would like to thank Y. Murakami, H. Nakao and H. Shiba 
for valuable discussion.
This work was partially supported by 
a Grant-in-Aid for Scientific Research 
from the Ministry of Education, Science, Sports and Culture, Japan.


\begin{thebibliography}{99}
\bibitem{Com1} Details of the quadrupole process will be discussed elsewhere.
\end{thebibliography}
\end{document}